\documentclass[preprint,12pt]{elsarticle}

\usepackage{amsmath,amsfonts}
\usepackage{algorithmic}
\usepackage{array}
\usepackage[tight,footnotesize]{subfigure}
\usepackage{textcomp}
\usepackage{stfloats}
\usepackage{url}
\usepackage{verbatim}
\usepackage{graphicx}
\def\BibTeX{{\rm B\kern-.05em{\sc i\kern-.025em b}\kern-.08em
    T\kern-.1667em\lower.7ex\hbox{E}\kern-.125emX}}
\usepackage{balance}
\usepackage{booktabs}% for \toprule, \midrule, and \bottomrule
\usepackage{caption}
\usepackage{makecell}
\usepackage{adjustbox}
\usepackage{wrapfig}
\usepackage[linesnumbered,ruled]{algorithm2e}
\usepackage{amssymb}
\usepackage{lineno}
\usepackage{float}
\usepackage{algorithmic}
\usepackage[tight,footnotesize]{subfigure}
\usepackage{booktabs}% for \toprule, \midrule, and \bottomrule
\usepackage{caption}
\usepackage{makecell}
\usepackage{adjustbox}
\usepackage{wrapfig}
\usepackage{xcolor}
\usepackage[linesnumbered,ruled]{algorithm2e}

\newcommand{\cmark}{\ding{51}}%
\newcommand{\xmark}{\ding{55}}%

\journal{Journal Name}

\begin{document}
%\pagenumbering{gobble} 
\sloppy
\setlength{\parskip}{0pt}

\begin{frontmatter}

\title{Train the Trainers --- An Agentic AI Framework for Peer-Based Mental Health Support in Battlefield Environments}

%\title{Designing, Developing, and Deploying Production-Grade Agentic AI Workflows}

%\title{Designing, Developing, and Deploying Production-Grade Agentic AI Workflows: A Practical Guide}

% % use optional labels to link authors explicitly to addresses:
\author[label2]{Atmaram Yarlagadda}
\ead{atmaram.yarlagadda.civ@health.mil}
\author[label1]{Eranga Bandara}
\ead{cmedawer@odu.edu}
\author[label1]{Ross Gore}
\ead{rgore@odu.edu}
\author[label4]{Anita H.\ Clayton}
\ead{AHC8V@uvahealth.org}
\author[label3]{Preston Samuel}
\ead{preston.l.samuel.mil@health.mil}
\author[label1]{Christopher K.\ Rhea}
\ead{crhea@odu.edu}
\author[label1]{Sachin Shetty}
\ead{sshetty@odu.edu}
\author[label1]{Ravi Mukkamala}
\ead{mukka@odu.edu}
\author[label5]{Xueping Liang}
\ead{xuliang@fiu.edu}
\author[label6]{Amin Hass}
\ead{amin.hassanzadeh@accenture.com}
\author[label7]{Abdul Rahman}
\ead{abdulrahman@deloitte.com}

\address[label2]{McDonald Army Health Center, Newport News, VA, USA}
\address[label1]{Old Dominion University, Norfolk, VA, USA}
\address[label3]{Blanchfield Army Community Hospital, Fort Campbell, KY, USA}
\address[label4]{Department of Psychiatry and Neurobehavioral Sciences, \\ University of Virginia School of Medicine, Charlottesville, VA, USA}
\address[label5]{Florida International University, FL, USA}
\address[label6]{Accenture Technology Labs, Arlington, VA, USA}
\address[label7]{Deloitte \& Touche LLP, USA}

\begin{abstract}

Modern military operations expose soldiers to sustained psychological stress, leading to acute reactions, post-traumatic stress symptoms, and other mental health issues. Although the U.S. Department of Defense offers evidence-based therapies, access to trained professionals in forward-deployed and contested environments is limited. As a result, soldiers with early-stage distress are often evacuated to rear medical facilities, delaying care, reducing readiness, and increasing long-term risks. 
This paper proposes a Train-the-Trainers framework in which soldiers who have completed therapy and returned to duty are trained as peer facilitators to provide first-line psychological support in operational settings. To scale and standardize this model under severe resource and connectivity constraints, we introduce an agentic AI–enabled platform that augments these recovered soldiers with specialized AI agents. The recovered soldier acts as a human supervisor, coordinating agents for symptom triage, guided peer-support interventions, operational constraint reasoning, training and simulation, and structured documentation for clinical escalation when needed.
To enable responsible and explainable decisions in high-stakes environments, the AI agents use a consortium of fine-tuned language models plus a reasoning-oriented model, applying consensus-based inference to generate recommendations. This approach reduces single-model bias, improves robustness, and provides transparent justifications while preserving human-in-the-loop control. The architecture functions in air-gapped and low-connectivity settings, maintaining human oversight and ethical safeguards.
A functional prototype was developed with the McDonald U.S. Army Health Center, Newport News, VA, USA. By combining peer-based intervention with consensus-driven agentic AI decision support, the framework seeks to cut response times, prevent symptom escalation, reduce unnecessary evacuations, and improve continuity of care. This work shows how agentic AI can serve as a force multiplier for mental health support in austere environments and identifies pathways for broader evaluation and deployment across defense and humanitarian operations.

\end{abstract}


\end{frontmatter}

\section{Introduction}

Modern military operations increasingly expose soldiers to prolonged psychological stress arising from combat engagement, asymmetric warfare, extended deployments, and morally injurious experiences. These conditions contribute to a wide spectrum of mental health challenges, including acute stress reactions, anxiety disorders, depression, post-traumatic stress symptoms, and cognitive impairment~\cite{military-stress-1}. Early identification and timely intervention are critical, as untreated or delayed psychological distress can escalate into severe long-term conditions, negatively affecting both individual well-being and overall force readiness.

The U.S. Department of Defense (DoD) has established evidence-based therapeutic programs delivered by trained mental health professionals, including psychologists, psychiatrists, and specialized therapists. However, the availability of such professionals is inherently limited in forward-deployed and contested environments~\cite{military-stress-2, proof-of-tbi}. Battlefield conditions, logistical constraints, operational tempo, and intermittent or absent connectivity often prevent timely access to professional assessment and care. As a result, soldiers experiencing early or moderate psychological distress are frequently evacuated to rear medical facilities or base hospitals, including those located outside the theater of operations. While clinically necessary in severe cases, such evacuations can delay intervention, disrupt unit cohesion, reduce operational effectiveness, and impose additional logistical and human costs~\cite{military-stress-3}.

To address these limitations, military mental health research has increasingly explored peer-based support models, recognizing that soldiers often trust and engage more readily with peers who share similar operational experiences~\cite{peer-based-tranning-1}. In this context, soldiers who have successfully undergone therapeutic interventions and returned to duty represent an underutilized resource. Their lived experience and targeted training enable them to recognize early warning signs, provide immediate peer support, and escalate to professional care when needed. This observation motivates a Train-the-Trainers approach, in which recovered soldiers are trained as peer facilitators capable of delivering first-line psychological support within the battlefield environment. Importantly, this model does not seek to replace professional clinicians but rather to augment existing care pathways by enabling early intervention and informed decision-making at the point of need.

Despite its promise, implementing a Train-the-Trainers model at scale presents several challenges. Peer facilitators must operate under high cognitive load, rapidly evolving operational constraints, and limited access to expert guidance~\cite{peer-based-tranning-2}. Ensuring consistency, safety, and ethical oversight across peer-led interventions is particularly difficult in austere environments. Without structured support, peer facilitators may struggle to assess symptom severity, select appropriate interventions, or determine when escalation is necessary~\cite{military-stress-2}.

Recent advances in agentic artificial intelligence (AI) offer a promising pathway to address these challenges~\cite{agentic-ai, agentic-ai-scientific}. Agentic AI systems consist of multiple specialized agents capable of perception, reasoning, decision support, and coordination, operating under human supervision. Rather than functioning as autonomous decision-makers, such systems are designed to augment human judgment, provide structured guidance, and manage complex workflows across constrained environments. Prior work has demonstrated the effectiveness of human-in-the-loop multi-agent architectures in domains requiring high reliability, adaptability, and accountability~\cite{agentic-workflow-practicle-guide}.  

In this paper, we propose an agentic AI–enabled Train-the-Trainers framework for battlefield mental health support.In the proposed system, a recovered soldier, trained as a peer facilitator, acts as a human supervisor. This supervisor orchestrates a suite of specialized AI agents that support symptom triage, guided peer interventions, operational constraint reasoning, training and simulation, and structured documentation~\cite{rovanima}. The platform is designed to operate in air-gapped or low-connectivity environments, making it suitable for deployment in forward and contested operational settings~\cite{air-gapped-2}. Human oversight is preserved at all times, with AI agents providing decision support rather than autonomous clinical judgment. To promote responsible, robust, and explainable decision-making in high-stakes military contexts, the AI agents are configured to operate over a consortium of fine-tuned language models combined with a reasoning-oriented model, using a consensus-driven inference mechanism to generate recommendations~\cite{responsible-ai, xai}. This approach mitigates single-model bias, improves reliability under uncertainty, and enables transparent justification of agent outputs, while maintaining strict human-in-the-loop control and supporting responsible and explainable AI principles~\cite{towards-rai-xai}. By positioning recovered soldiers as peer facilitators, supported by consensus-driven agentic AI, the proposed framework leverages both lived experience and structured decision support. This combined approach addresses the challenges of timely assessment, peer-level intervention, and informed escalation in austere and resource-constrained environments.The primary contributions of this paper are fourfold:

\begin{enumerate}
    \item The formulation of a Train-the-Trainers model tailored to battlefield mental health support
    \item The design of an agentic AI architecture that augments recovered soldiers as peer facilitators under operational constraints
    \item A conceptual workflow illustrating how human-AI collaboration can reduce response latency, prevent symptom escalation, and improve continuity of care.
    \item The implementation of a functional prototype of the proposed platform, developed in collaboration with the McDonald U.S. Army Health Center, Newport News, VA, USA.
\end{enumerate}

The remainder of this paper is organized as follows. Section~2 provides background on the key concepts underpinning this work, including large language models, agentic AI, and responsible and explainable AI. Section~3 describes the entities involved in the proposed Train-the-Trainers framework, including human roles and specialized AI agents. Section~4 details the operational functionality of the system, focusing on the peer-support workflow and human–AI interaction model. Section~5 presents the implementation and evaluation of the proposed platform. Section~6 reviews related work in military mental health support, peer-based interventions, and agentic AI systems. Finally, Section~7 concludes the paper and outlines directions for future research.

\section{Background}

This section introduces the key technical concepts underlying the proposed Train-the-Trainers framework. We briefly review large language models, reasoning-oriented language models, fine-tuning approaches, agentic AI systems, and responsible and explainable AI principles, which together form the foundation of the proposed architecture.

\subsection{Large Language Models (LLMs)}

Large Language Models (LLMs) are deep neural networks trained on large-scale textual corpora to learn statistical representations of language~\cite{llm}. Through pre-training on diverse data, LLMs acquire the ability to generate coherent text, perform question answering, summarize information, and reason over complex natural language inputs. Recent advances have demonstrated that LLMs can be adapted to domain-specific tasks through fine-tuning, enabling their use in decision-support applications across healthcare, defense, and other high-stakes domains~\cite{gpt-llm}.

In the context of this work, LLMs serve as the core reasoning engines that process structured and unstructured assessment inputs, generate interpretable outputs, and support human decision-making rather than autonomous action.

\subsection{Reasoning-Oriented Language Models}

Reasoning-oriented language models extend standard LLM capabilities by emphasizing multi-step inference, comparison of alternatives, and explicit justification of outputs. These models are designed to evaluate multiple candidate responses, assess coherence and contextual relevance, and synthesize a final output that reflects structured reasoning rather than pattern completion alone~\cite{reasoning-llms, llm-reasoning}.

Within the proposed framework, a reasoning-oriented LLM acts as a supervisory component that synthesizes outputs from multiple fine-tuned LLMs. This consensus-driven reasoning approach improves robustness, reduces single-model bias, and enables the presentation of transparent rationales to support explainable and accountable decision support.

\subsection{LLM Fine-Tuning}

Fine-tuning is the process of adapting a pre-trained LLM to a specific task or domain by training it further on curated datasets. Techniques such as Low-Rank Adapters (LoRA) and quantization-aware fine-tuning enable efficient adaptation while reducing computational and memory requirements, making fine-tuned models suitable for deployment in resource-constrained environments~\cite{llm-finetune, proof-of-tbi}.

In this work, fine-tuning is used to specialize LLMs for structured psychological assessment tasks. Fine-tuning is performed offline or through secure update pipelines, ensuring that system behavior remains stable during live operation and compatible with air-gapped deployment requirements.

\subsection{AI Agents and Agentic AI}

AI agents represent a paradigm in which LLMs are embedded within autonomous or semi-autonomous software entities capable of interacting with their environment, invoking tools, maintaining state, and executing multi-step workflows. An AI agent can autonomously perform an interaction loop by constructing prompts, invoking language models, interpreting responses, and triggering subsequent actions without direct human intervention~\cite{agentic-ai, agentic-workflow-practicle-guide}.

\begin{figure}[H]
\centering
\includegraphics[width=5.3in]{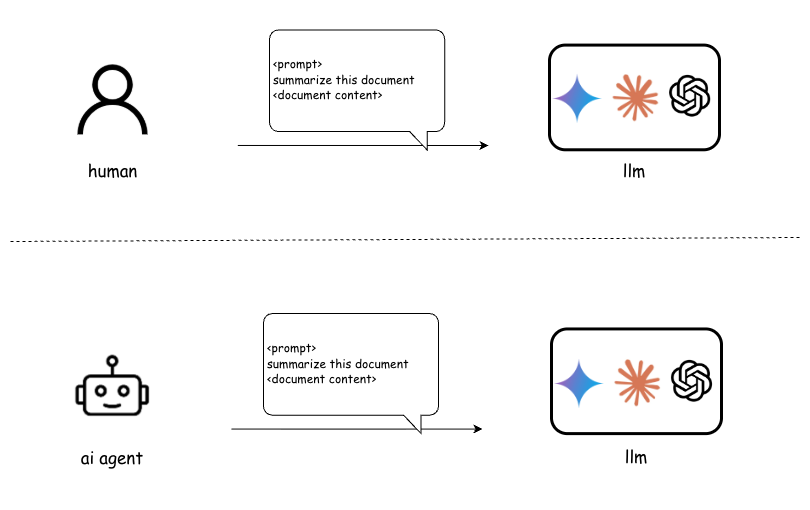}
\vspace{-0.1in}
\caption{Comparison between direct human--LLM interaction and agentic AI--LLM interaction. In the top panel, a human user directly constructs and submits a prompt to a language model, manually interpreting the response. In the bottom panel, an AI agent autonomously manages the interaction loop by constructing prompts, invoking the language model, interpreting responses, and determining subsequent actions without direct human intervention. This agent-mediated workflow enables iterative reasoning, tool integration, and task automation, forming the basis of agentic AI systems used for structured decision support and multi-step workflows.}
\label{ai-agent}
\end{figure}

Figure~\ref{ai-agent} contrasts traditional human--LLM interaction with autonomous AI agent--LLM interaction. In essence, AI agents are software programs that leverage LLMs in combination with tools, application programming interfaces (APIs), and external context to execute tasks iteratively. Agentic AI systems extend this concept by coordinating multiple specialized agents under a shared orchestration framework.

In the proposed Train-the-Trainers framework, agentic AI is employed in a human-supervised configuration, where AI agents provide structured decision support while authority and execution remain under human control.

\subsection{Responsible and Explainable AI}

Responsible and Explainable AI principles emphasize transparency, accountability, fairness, and human oversight in AI system design. In high-stakes domains such as military operations and mental health support, these principles are critical for ensuring trust, safety, and ethical deployment~\cite{towards-rai-xai}.

Explainable AI focuses on providing interpretable outputs and rationales that allow human operators to understand and evaluate system recommendations. Responsible AI further requires clear operational boundaries, avoidance of autonomous decision-making in sensitive contexts, and mechanisms for auditability and oversight~\cite{responsible-gen-ai, llm-explainability}.

The proposed framework operationalizes these principles through strict human-in-the-loop control, consensus-driven reasoning across multiple models, structured logging of agent interactions, and offline learning mechanisms. This design ensures that agentic AI augments human capability without displacing human judgment or authority.

\section{Entities in the System}

This section describes the primary human and artificial entities involved in the proposed Train-the-Trainers framework. The system is designed to preserve clear boundaries of authority, ensuring that all decision-making remains under human supervision while AI agents provide structured decision support in austere and resource-constrained environments.

\subsection{Human Entities}

\subsubsection{Peer Trainer (Recovered Soldier)}

The peer trainer is a recovered soldier who has successfully completed therapeutic intervention and returned to active duty. In the proposed framework, the peer trainer serves as the \emph{human supervisor} of the agentic AI platform and is responsible for initiating, overseeing, and executing all interactions within the system~\cite{rovanima}. The peer trainer is not a licensed clinician and does not perform diagnosis or treatment; instead, they provide first-line peer support informed by training and augmented by AI-based decision support.

The peer trainer retains full authority over all actions, including whether to engage AI agents, how to interpret their recommendations, and when to escalate a situation to professional clinical services. By combining lived operational experience with structured guidance from AI agents, the peer trainer is positioned to deliver timely and context-aware support in forward-deployed environments while maintaining ethical and procedural boundaries~\cite{agent-survey}.

\subsubsection{Soldier Receiving Support}

The soldier receiving support is an active-duty service member experiencing psychological distress during deployment or operational activities. Interactions between the peer trainer and the supported soldier occur within a peer-to-peer context, emphasizing trust, shared experience, and reduced stigma. Participation is voluntary, and the system is designed to respect privacy while minimizing unnecessary data collection~\cite{peer-based-tranning-1}.

The supported soldier does not interact directly with the AI agents. All AI-mediated guidance is filtered through the peer trainer, ensuring that technology augments human interaction rather than replacing it. This design reinforces human-centered care and reduces the risk of inappropriate reliance on automated systems~\cite{agentic-workflow-practicle-guide}.

\subsection{AI Agents}

\begin{figure}[H]
\centering{}
\includegraphics[width=5.2in]{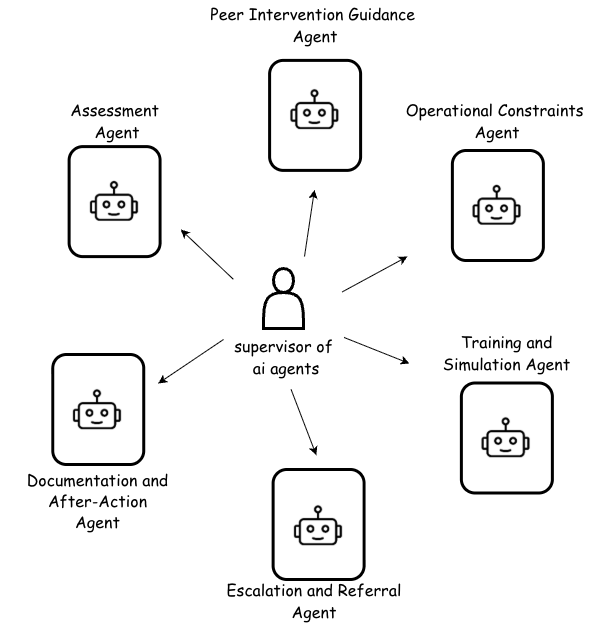}
\DeclareGraphicsExtensions.
\caption{Human-supervised agentic AI architecture for the Train-the-Trainers framework. A recovered soldier acting as a peer trainer serves as the central human supervisor, coordinating multiple specialized AI agents that provide advisory decision support for assessment, peer intervention guidance, operational constraint reasoning, escalation and referral, documentation and after-action reporting, and training and simulation. All agent interactions are advisory in nature, with authority and execution remaining strictly under human control.}
\label{train-agents}
\end{figure}

The agentic AI platform comprises multiple specialized AI agents, each responsible for a narrowly scoped function. All agents operate under strict human-in-the-loop control and do not perform autonomous clinical decision-making, as illustrated in Figure~\ref{train-agents}. Recommendations are generated using a consensus-driven inference process over a consortium of fine-tuned language models in combination with a reasoning-oriented model, supporting robustness, transparency, and explainability in high-stakes operational environments~\cite{towards-rai-xai, agentsway}.

\subsubsection{Assessment Agent}

The Assessment Agent supports structured, non-clinical evaluation of observable psychological and behavioral indicators during peer-support interactions. It assists the peer trainer in systematically checking symptoms, contextual signals, and functional impairments without performing a medical diagnosis. The agent operationalizes an assessment framework that captures early indicators of psychological distress, such as acute stress responses, emotional dysregulation, cognitive overload, sleep disruption, and behavioral withdrawal~\cite{psychiatric-disorders}.

The assessment process is guided by a structured framework that we previously developed for systematic symptom and context evaluation. Within the Train-the-Trainers system, this framework is embedded as an assessment logic layer that translates observed signals into interpretable assessment outputs, including severity bands, risk flags, and uncertainty indicators. The agent presents these outputs to the peer trainer as decision-support artifacts, enabling consistent and repeatable assessment across different operational contexts while explicitly avoiding clinical labeling or diagnosis~\cite{deep-psychiatric,med-palm}.

By standardizing early assessment, the Assessment Agent reduces subjective variability, supports early identification of concerning patterns, and provides a transparent basis for subsequent intervention or escalation decisions.

\subsubsection{Peer Intervention Guidance Agent}

The Peer Intervention Guidance Agent provides real-time decision support during peer-led interactions following assessment. Based on assessment outputs and ongoing observations, the agent suggests appropriate non-clinical stabilization and support techniques, including grounding exercises, breathing regulation, short rest-cycle interventions, and structured peer communication strategies~\cite{metal-health-challanges, llm-ddx}.

Recommendations adapt dynamically as the situation evolves, but remain strictly advisory. The agent does not prescribe actions or override human judgment; instead, it supports the peer trainer by reducing cognitive load during high-stress interactions and ensuring alignment with established peer-support practices.

\subsubsection{Operational Constraints Agent}

The Operational Constraints Agent reasons over contextual factors specific to battlefield and forward-deployed environments. These factors include mission posture, location, time sensitivity, unit availability, communication constraints, and access to medical or evacuation resources~\cite{drhouse}.

By integrating operational context with assessment and intervention information, the agent assists the peer trainer in identifying feasible courses of action that balance mental health support needs with mission requirements. This ensures that recommended interventions and escalation pathways remain practical, actionable, and aligned with operational realities~\cite{mental-health-ml}.

\subsubsection{Escalation and Referral Agent}

The Escalation and Referral Agent supports decision-making when peer-level intervention is insufficient or when assessment outputs indicate elevated risk. It assists the peer trainer in recognizing escalation thresholds and identifying appropriate referral pathways, such as on-site medical personnel, remote consultation when available, or evacuation to rear medical facilities~\cite{psychiatric-disorders}.

The agent generates structured handoff summaries that consolidate assessment findings, observed behaviors, actions taken, and contextual constraints. These summaries are designed to improve continuity of care and reduce information loss during transitions to professional services. Final escalation decisions remain entirely under the authority of the peer trainer.

\subsubsection{Documentation and After-Action Agent}

The Documentation and After-Action Agent assists in capturing minimal, structured records of peer-support interactions. It summarizes key assessment outputs, interventions applied, decisions made, and observed outcomes in a standardized format suitable for later review by authorized personnel~\cite{med-palm}.

The agent is designed to minimize administrative burden and cognitive overhead while supporting accountability, continuity of care, and after-action analysis. All documentation is privacy-preserving and avoids unnecessary personal or identifying information.

\subsubsection{Training and Simulation Agent}

The Training and Simulation Agent supports both initial preparation and ongoing skill maintenance for peer trainers. It provides scenario-based training exercises and simulated interactions that reflect common battlefield psychological stress situations, including ambiguous, time-pressured, and resource-constrained cases~\cite{dsm5-llm}.

Using feedback from assessment, intervention, and escalation outcomes, the agent offers structured feedback to peer trainers on communication effectiveness, assessment consistency, and decision-making quality. This enables continuous skill development without requiring constant access to human instructors~\cite{CDSS}.

\subsubsection{Fine-Tuning and Continuous Learning Agent}

The Fine-Tuning and Continuous Learning Agent enables adaptive improvement of the agentic AI platform over time. It aggregates inputs and outputs generated by other AI agents, including assessment results, intervention guidance, operational reasoning, escalation summaries, and training interactions, along with human feedback provided by peer trainers. These data are processed in a privacy-preserving and non-identifying manner~\cite{llm-finetune}.

Based on the aggregated information, the agent supports offline or securely managed fine-tuning of the language models that form the system’s model consortium. Updated models are incorporated into subsequent system iterations, enabling a continuous learning approach that improves robustness, contextual relevance, and decision-support quality while preserving human oversight and system stability~\cite{mistral-fine-tune, devsec-gpt}.

\section{Functionality of the System}

This section describes how the proposed Train-the-Trainers framework operates in practice. Rather than reiterating individual agent roles, the focus here is on the end-to-end operational workflow, the lifecycle of peer trainers, and the manner in which agentic AI supports human decision-making throughout training, deployment, and peer-support interactions in battlefield environments.

\subsection{Lifecycle of Peer Trainers}

Soldiers experiencing psychological distress during deployment are initially removed from active battlefield duties and referred to rear medical facilities or base hospitals, where they receive appropriate professional therapeutic interventions conducted by qualified mental health professionals. These interventions follow established clinical protocols and aim to support recovery, stabilization, and functional readiness~\cite{peer-based-tranning-2}.

Upon successful completion of therapy and clearance for return to duty, a subset of recovered soldiers may voluntarily enroll in additional preparation programs to serve as peer trainers. These programs focus on recognizing early indicators of psychological distress, applying non-clinical peer-support techniques, understanding escalation pathways, and operating within defined ethical and procedural boundaries. Scenario-based exercises and simulated interactions are used to reinforce decision-making and communication skills.

Soldiers who complete these requirements graduate as peer trainers and are equipped with the agentic AI–enabled platform prior to redeployment.

\subsection{Peer Support Interaction Workflow in Battlefield Environments}

Once deployed, peer trainers return to forward-deployed or contested operational environments as part of their assigned units. In addition to their primary operational responsibilities, they serve as first-line peer-support facilitators within their units. When a fellow soldier exhibits signs of psychological distress, the peer trainer initiates a peer-support interaction and engages the agentic AI platform as needed~\cite{peer-based-tranning-1}.

The peer-support workflow proceeds through a sequence of stages. First, the peer trainer conducts an initial engagement with the supported soldier, relying on direct observation and peer communication. The system then provides structured assessment outputs that help the peer trainer evaluate observable symptoms, contextual factors, and functional impacts in a consistent and non-clinical manner~\cite{military-stress-3}. Based on these assessment outputs, the peer trainer may apply appropriate peer-level stabilization and support techniques, such as grounding, breathing regulation, or short rest-cycle interventions.

Throughout the interaction, the system incorporates operational context, including mission posture, time constraints, and available resources, to ensure that recommended actions remain feasible in the battlefield environment. At defined decision points, the peer trainer determines whether peer-level intervention is sufficient or whether escalation to professional medical services is required. If escalation is necessary, the system supports structured handoff by consolidating relevant observations and actions into a concise summary for medical personnel~\cite{military-stress-2}.

All interactions remain human-centered: the peer trainer is the sole point of contact with the supported soldier, and all system outputs are advisory and subject to human judgment.

\subsection{Human--AI Interaction Model}

The proposed framework adopts a strict human-in-the-loop interaction model. The agentic AI platform does not act autonomously or initiate actions independently. Instead, it provides structured decision support at specific stages of the peer-support workflow, presenting recommendations, risk indicators, and contextual considerations to the peer trainer~\cite{agentic-ai-rise, agentsway}.

In cases where multiple perspectives are available, system outputs are derived through a consensus-driven reasoning process that synthesizes inputs from multiple fine-tuned language models and a reasoning-oriented model. Recommendations are accompanied by interpretable rationales to support transparency and informed human decision-making. The peer trainer retains full authority to accept, modify, or disregard any system recommendation~\cite{rovanima}.

\subsection{Continuous Learning and System Evolution}

System learning occurs outside of live peer-support interactions to preserve stability, safety, and operational reliability. Following deployments or structured training cycles, the system aggregates structured outputs generated during assessments, peer-level interventions, escalation decisions, documentation activities, and training exercises, together with human feedback provided by peer trainers. All collected data are processed in a privacy-preserving and non-identifying manner~\cite{llm-finetune, llama-recipe, lora}.

\begin{figure}[H]
\centering{}
\includegraphics[width=5.4in]{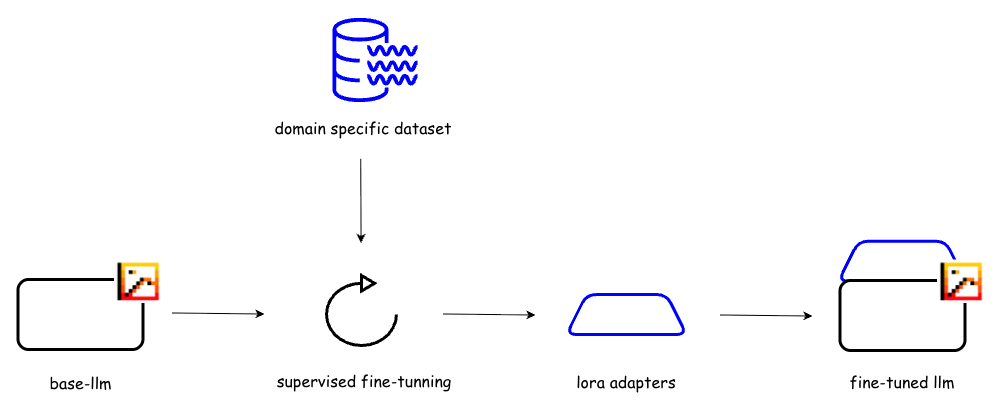}
\DeclareGraphicsExtensions.
\caption{Overview of the supervised fine-tuning pipeline used to adapt a base LLM to domain-specific tasks. A curated, domain-specific dataset is used to perform supervised fine-tuning of a base LLM, with Low-Rank Adapter (LoRA) modules applied to enable parameter-efficient training. The resulting fine-tuned model retains the general linguistic capabilities of the base model while incorporating domain-specific knowledge, supporting efficient deployment and updates in resource-constrained and air-gapped environments.}
\label{llm-fine-tune}
\end{figure}

The aggregated information supports offline or securely managed fine-tuning of the language models that form the system’s model consortium. Fine-tuning is performed between operational cycles and does not modify system behavior during active use. Updated models are incorporated into subsequent system iterations, enabling continuous improvement in contextual awareness, robustness, and decision-support quality while maintaining compatibility with air-gapped and low-connectivity environments. Figure~\ref{llm-fine-tune} illustrates the fine-tuning workflow and deployment process used to update language models safely across iterations~\cite{wedagpt, mistral-fine-tune}.

\subsection{Operational Boundaries and Safety Considerations}

The proposed system is explicitly designed as a decision-support platform and does not perform clinical diagnosis, treatment, or autonomous intervention. All recommendations generated by the agentic AI platform are advisory in nature, and all interpretation, decision-making, and execution remain under direct human control~\cite{agentic-ai-scientific, agentsway}. The peer trainer serves as the sole authority responsible for actions taken during peer-support interactions.

\begin{figure}[H]
\centering{}
\includegraphics[width=5.2in]{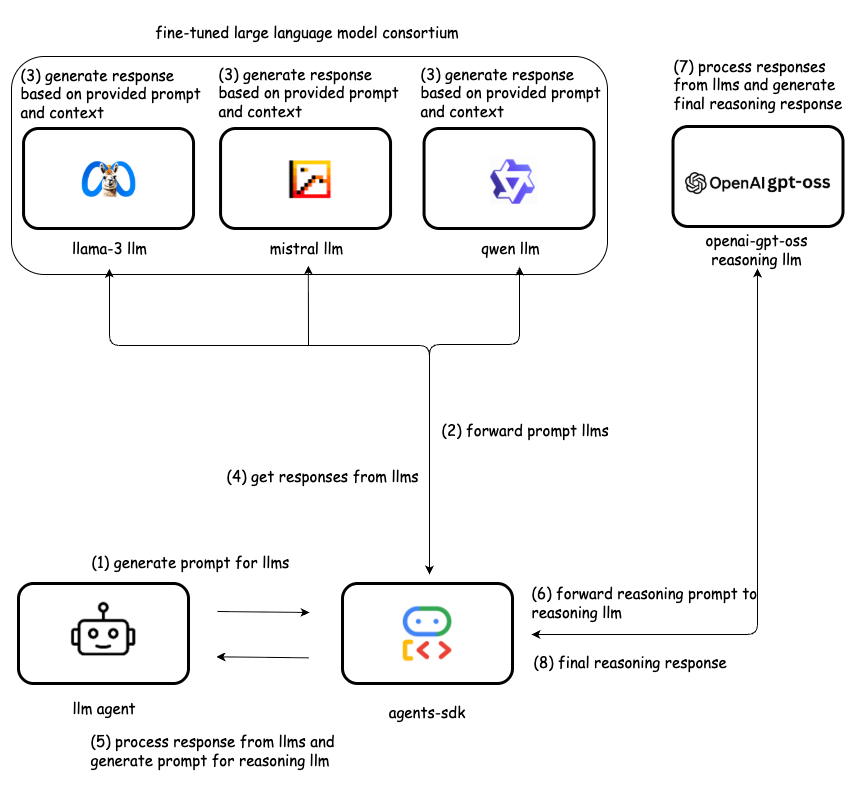}
\DeclareGraphicsExtensions.
\caption{Consensus-driven LLM consortium and reasoning workflow used by the agentic AI platform. An LLM agent generates a task-specific prompt, which is forwarded via the agent orchestration layer to multiple fine-tuned LLMs (e.g., Llama-3, Mistral, and Qwen). Each model independently generates a response based on the same prompt and contextual inputs. The resulting outputs are collected and processed by the orchestration layer, which constructs a structured reasoning prompt for a dedicated reasoning-oriented LLM (GPT-OSS). The reasoning LLM evaluates, reconciles, and synthesizes the multiple model outputs to produce a final consolidated response accompanied by an interpretable rationale. This consensus-driven process improves robustness, mitigates single-model bias, and supports explainable, human-supervised decision support.}
\label{llm-consortium}
\end{figure}

To support responsible and explainable AI use in high-stakes environments, the system employs a consensus-driven reasoning architecture in which recommendations are synthesized across multiple fine-tuned language models and a reasoning-oriented model. This design promotes robustness, reduces single-model bias, and enables the presentation of interpretable rationales alongside recommendations~\cite{towards-rai-xai}. The system operates locally without reliance on continuous network connectivity, making it suitable for deployment in austere and contested environments. Figure~\ref{llm-consortium} illustrates the consensus-driven LLM consortium and reasoning flow that underpin responsible, transparent, and human-supervised decision support.

\section{Implementation and Evaluation}

This section presents the implementation and evaluation of the proposed Train-the-Trainers platform. We describe the realization of a functional prototype designed to demonstrate the feasibility of deploying a human-supervised, agentic AI system for peer-based mental health support in forward-deployed and resource-constrained environments~\cite{rovanima}. The section then outlines the primary assessment use case implemented within the platform and defines the scope of the evaluation conducted to assess the technical performance of the underlying agentic AI components~\cite{agentic-ai-opptunities}.

\subsection{Implementation Overview}

The Train-the-Trainers framework was implemented as a functional prototype to show the feasibility of a human-supervised, agentic AI platform for peer-based mental health support in forward-deployed, resource-constrained settings. It supports recovered soldiers serving as peer trainers by enabling structured assessment, intervention guidance, escalation, documentation, training and simulation, and continuous learning without requiring constant network connectivity.

The system integrates multiple specialized AI agents, each responsible for a distinct support function, including assessment, peer intervention guidance, operational constraint reasoning, escalation and referral support, documentation and after-action reporting, training and simulation, and fine-tuning with continuous learning~\cite{proof-of-tbi}. All agents operate under a strict human-in-the-loop orchestration model, with the peer trainer serving as the sole authority responsible for interpreting recommendations and executing actions.

Each AI agent was implemented using the OpenAI Agents SDK~\cite{agentic-ai, deep-stride}, which provides modular primitives for defining agent roles, behaviors, and reasoning strategies. To enable interoperability and structured agent-to-agent communication, all agent functionalities were exposed through a Model Context Protocol (MCP)~\cite{mcp2, mcp1, mcc} server interface. This design allows agents to exchange structured assessment outputs, contextual signals, and recommendation artifacts through standardized endpoints while maintaining clear boundaries of responsibility.

For human interaction, the prototype employs an MCP-compatible interface tool that allows peer trainers to interact with individual agents, review consensus-driven recommendations, and control system behavior without requiring direct code-level intervention~\cite{agentsway, agentic-workflow-practicle-guide}. This interface supports transparent orchestration, enabling peer trainers to supervise multi-agent interactions, request specific support functions, and review structured rationales accompanying agent outputs.

To support deployment in contested and austere environments, the prototype was designed to operate in air-gapped or low-connectivity settings~\cite{air-gapped-1}. All core functionalities, including agent execution, reasoning, and decision support, run locally. Network connectivity is required only for periodic system updates or offline fine-tuning workflows, which are performed outside active operational contexts.

LLM fine-tuning was conducted using the Unsloth library~\cite{llamafactory-unsloth} on GPU-enabled environments equipped with NVIDIA A100 accelerators~\cite{a100-gpu, google-tpu}. To ensure efficient training and compatibility with constrained hardware, Low-Rank Adapters (LoRA)~\cite{lora} combined with 4-bit quantization (QLoRA)~\cite{qlora, vindsec-llama} were employed. The system utilizes a consortium of open-source language models, including Llama-3, Pixtral, and Qwen, each fine-tuned for specific support functions such as structured assessment, communication guidance, and operational reasoning~\cite{llama-3, pixtral, qwen2}.

A dedicated reasoning-oriented language model (GPT-OSS) serves as the supervisory component within the consortium, responsible for synthesizing and validating outputs generated by individual models~\cite{gpt-oss, reasoning-llms, llm-reasoning}. This consensus-driven reasoning process improves robustness, reduces single-model bias, and enables the generation of interpretable rationales to support explainable and responsible decision support.

Overall, the prototype demonstrates that the proposed agentic AI architecture can be effectively implemented using MCP-based orchestration and fine-tuned language models, operating reliably in air-gapped and resource-constrained environments~\cite{mcc}. These results indicate the feasibility of deploying the Train-the-Trainers framework as a scalable and adaptable decision-support system for peer-based mental health support in military and humanitarian contexts.

\subsection{Assessment Agent Use Case and Workflow}

The primary use case evaluated in this work is a psychiatric assessment support workflow implemented as the Assessment Agent within the agentic AI platform. The Assessment Agent is designed to support structured, non-clinical evaluation of observable psychological and behavioral indicators, enabling peer trainers to perform consistent early assessment without performing medical diagnosis.

The workflow follows a consensus-driven reasoning approach. Structured assessment inputs derived from observable symptoms, contextual information, and interaction cues are processed by a consortium of fine-tuned LLMs, each specialized to capture different aspects of assessment reasoning~\cite{towards-rai-xai}. These models generate independent assessment outputs reflecting diverse reasoning perspectives. A reasoning-oriented LLM (GPT-OSS) then evaluates, compares, and synthesizes these outputs to produce a consolidated assessment representation accompanied by an interpretable rationale.

This design aligns with the assessment framework discussed in related work and reflects realistic operational usage, where peer trainers rely on structured decision support rather than autonomous classification or diagnosis. The Assessment Agent operates entirely in an advisory capacity, with all interpretations and decisions remaining under human control.

\subsection{Evaluation of LLM Fine-Tuning}

To evaluate the fine-tuning process used to support the Assessment Agent, we fine-tuned the Qwen2.5-7B-Instruct model~\cite{qwen2} on a curated psychiatric assessment dataset. The dataset consists of 500 synthetic psychiatrist–patient conversations spanning five DSM-5 reference categories: Major Depressive Disorder (DSM-5 296.2x), Generalized Anxiety Disorder (DSM-5 300.02), Panic Disorder (DSM-5 300.01), Post-Traumatic Stress Disorder (DSM-5 309.81), and Bipolar I Disorder (DSM-5 296.4x)~\cite{dsm-5, dsm-5-criteria}. These categories were used as structured reference labels to evaluate assessment alignment rather than to enable autonomous diagnosis.

\begin{figure}[H]
\centering{}
\includegraphics[width=5.3in]{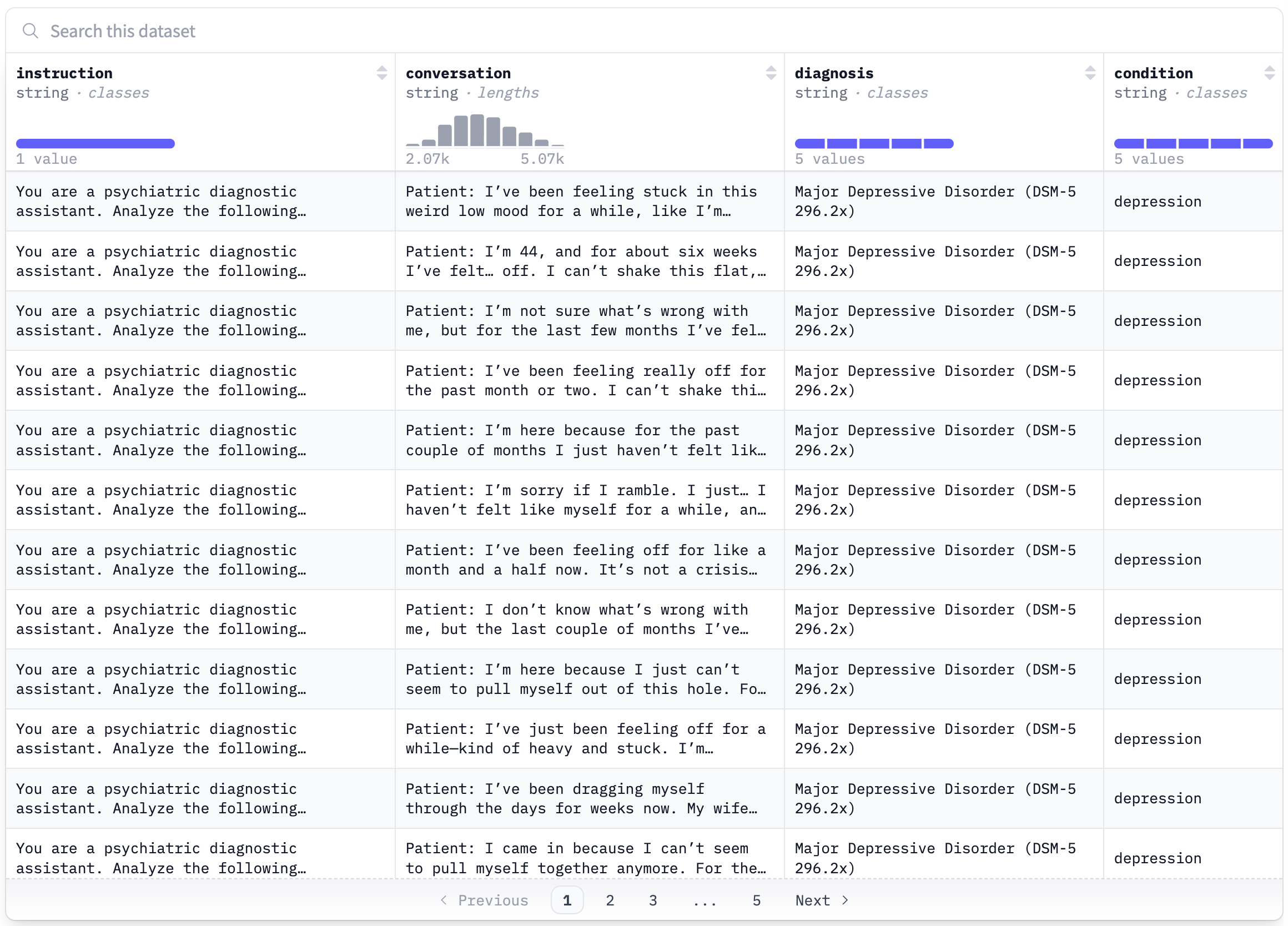}
\DeclareGraphicsExtensions.
\caption{Structure of the psychiatric assessment fine-tuning dataset. Each sample contains an instruction prompt, a psychiatrist-patient conversation (2.07k--5.07k characters), a DSM-5 diagnosis label, and a condition category. The dataset was constructed from publicly available data and synthetically generated conversations covering five psychiatric conditions.}
\label{fig:mental-dataset}
\end{figure}

The fine-tuning dataset was constructed using a combination of publicly available psychiatric dialogue data and synthetically generated conversations. Synthetic samples were generated using large language models guided by established diagnostic criteria to ensure realistic symptom expression, contextual coherence, and adherence to DSM-5 assessment patterns~\cite{dsm-5-traumatic-stress, dsm5-llm}. Figure~\ref{fig:mental-dataset} illustrates the structure of the fine-tuning dataset.

Each training sample consists of four components: (1) \textit{instruction}, a standardized system prompt directing the model to perform structured assessment reasoning; (2) \textit{conversation}, a full psychiatrist–patient dialogue ranging from 2{,}070 to 5{,}070 characters and capturing symptom presentation, patient history, and assessment-oriented questioning; (3) \textit{diagnosis}, representing the reference DSM-5 category used for supervised learning; and (4) \textit{condition}, a categorical label identifying the corresponding assessment class. Table~\ref{tab:mental-dataset} summarizes the dataset composition and class distribution~\cite{dsm5-llm}.

\begin{table}[H]
\centering
\caption{Psychiatric Assessment Dataset Statistics}
\label{tab:mental-dataset}
\begin{tabular}{lcc}
\toprule
\textbf{Category} & \textbf{Count} & \textbf{Percentage} \\
\midrule
Total Samples & 500 & 100.0\% \\
\midrule
Training Set & 400 & 80.0\% \\
Validation Set & 50 & 10.0\% \\
Test Set & 50 & 10.0\% \\
\midrule
\multicolumn{3}{l}{\textit{Training Set Distribution by Condition:}} \\
\quad Depression & 82 & 20.5\% \\
\quad Panic Disorder & 83 & 20.8\% \\
\quad PTSD & 83 & 20.8\% \\
\quad Bipolar Disorder & 79 & 19.8\% \\
\quad Anxiety Disorder & 73 & 18.3\% \\
\bottomrule
\end{tabular}
\end{table}

Fine-tuning was performed using the Unsloth library on Google Colab with an NVIDIA A100-SXM4-40GB GPU~\cite{llamafactory-unsloth, a100-gpu}. The base model was loaded in 4-bit quantized format to optimize memory utilization. We employed Low-Rank Adaptation (LoRA)~\cite{lora} with rank $r=32$ and $\alpha=32$, targeting the attention and MLP projection layers (q\_proj, k\_proj, v\_proj, o\_proj, gate\_proj, up\_proj, down\_proj). This configuration resulted in 40,370,176 trainable parameters, representing only 0.53\% of the total 7.66 billion model parameters. Table~\ref{tab:mental-training-config} details the complete training configuration.

\begin{table}[H]
\centering
\caption{Fine-Tuning Configuration for Psychiatric Assessment Model}
\label{tab:mental-training-config}
\begin{tabular}{ll}
\toprule
\textbf{Parameter} & \textbf{Value} \\
\midrule
\multicolumn{2}{l}{\textit{Model Configuration:}} \\
Base Model & Qwen2.5-7B-Instruct (4-bit) \\
Maximum Sequence Length & 4,096 tokens \\
Precision & BFloat16 \\
\midrule
\multicolumn{2}{l}{\textit{Training Configuration:}} \\
Per-device Batch Size & 2 \\
Gradient Accumulation Steps & 4 \\
Effective Batch Size & 8 \\
Maximum Training Steps & 200 ($\approx$4 epochs) \\
Learning Rate & $1 \times 10^{-4}$ \\
Warmup Steps & 20 \\
Learning Rate Scheduler & Linear decay \\
Optimizer & AdamW (8-bit) \\
Weight Decay & 0.01 \\
Early Stopping Patience & 10 evaluations \\
\midrule
\multicolumn{2}{l}{\textit{LoRA Configuration:}} \\
LoRA Rank ($r$) & 32 \\
LoRA Alpha ($\alpha$) & 32 \\
LoRA Dropout & 0 \\
Target Modules & q, k, v, o, gate, up, down proj \\
Trainable Parameters & 40,370,176 (0.53\%) \\
Total Model Parameters & 7,655,986,688 \\
\bottomrule
\end{tabular}
\end{table}

\subsubsection{Training Convergence Analysis}

\begin{figure}[H]
\centering{}
\includegraphics[width=5.2in]{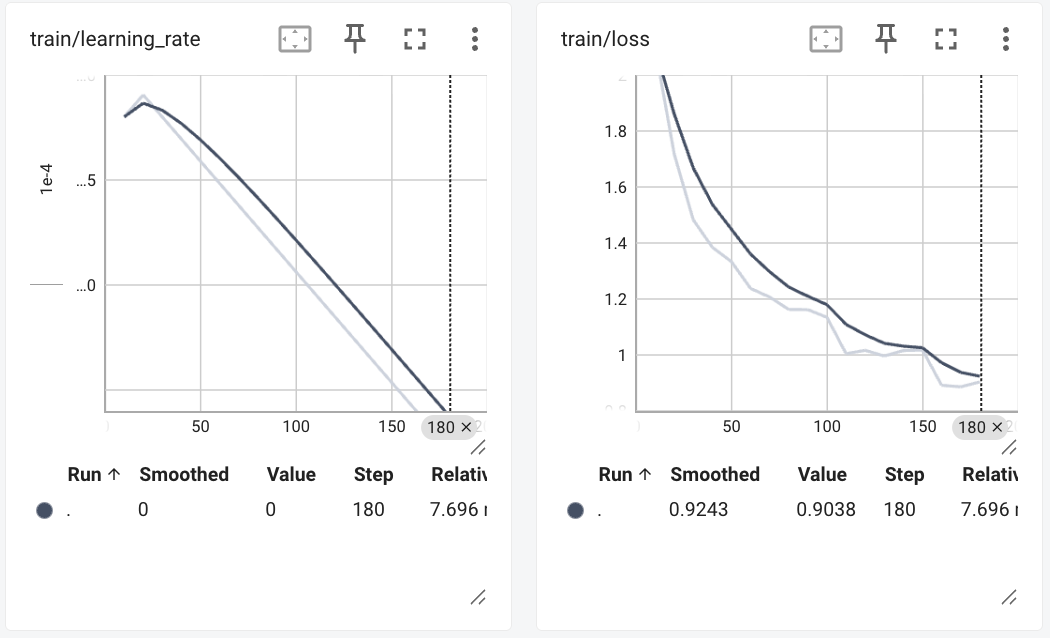}
\DeclareGraphicsExtensions.
\caption{Training metrics during fine-tuning: learning rate schedule (left) showing warmup to $1 \times 10^{-4}$ followed by linear decay, and training loss (right) decreasing from 1.9 to 0.90 over 180 steps, indicating effective learning of the diagnostic classification task.}
\label{fig:mental-train-loss}
\end{figure}

Training and validation loss curves were monitored via TensorBoard to assess convergence behavior and detect potential overfitting. Figure~\ref{fig:mental-train-loss} illustrates the training loss progression, which decreased from an initial value of approximately 1.9 to 0.90 over 180 training steps, demonstrating consistent learning without significant instability.

\begin{figure}[H]
\centering{}
\includegraphics[width=5.2in]{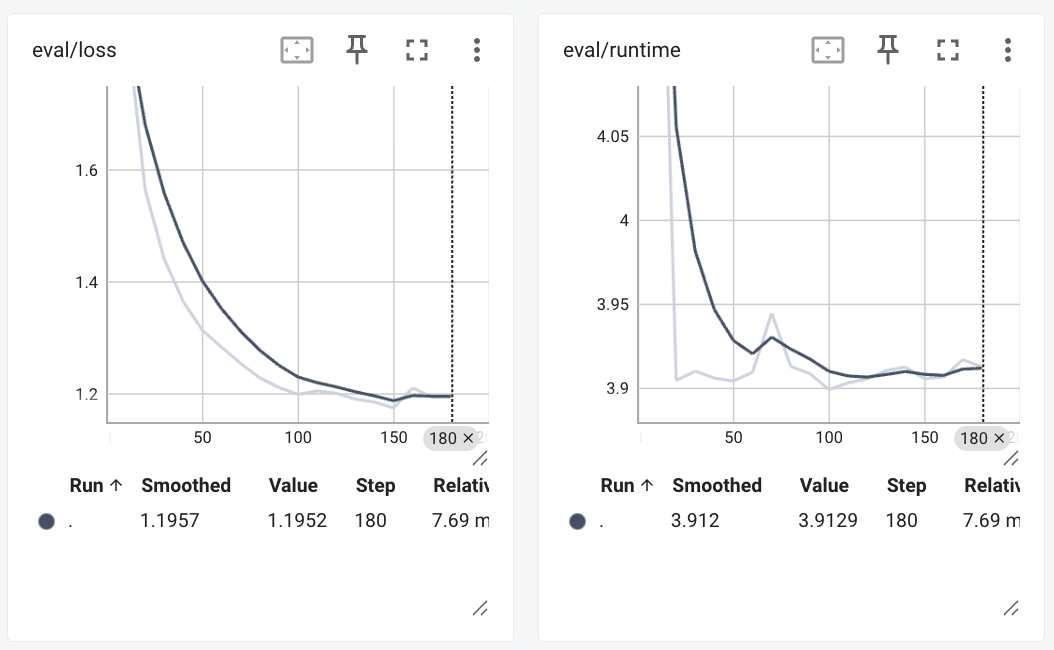}
\DeclareGraphicsExtensions.
\caption{Validation metrics during fine-tuning: evaluation loss (left) decreasing from 1.8 to 1.195, closely following the training loss trajectory and indicating good generalization without overfitting, and evaluation runtime (right) stabilizing at approximately 3.91 seconds.}
\label{fig:mental-eval-loss}
\end{figure}

The validation loss, shown in Figure~\ref{fig:mental-eval-loss}, decreased from approximately 1.8 to 1.195, closely tracking the training loss without significant divergence. This parallel decrease indicates that the model generalized well to unseen data without overfitting to the training set.

\begin{figure}[H]
\centering{}
\includegraphics[width=5.2in]{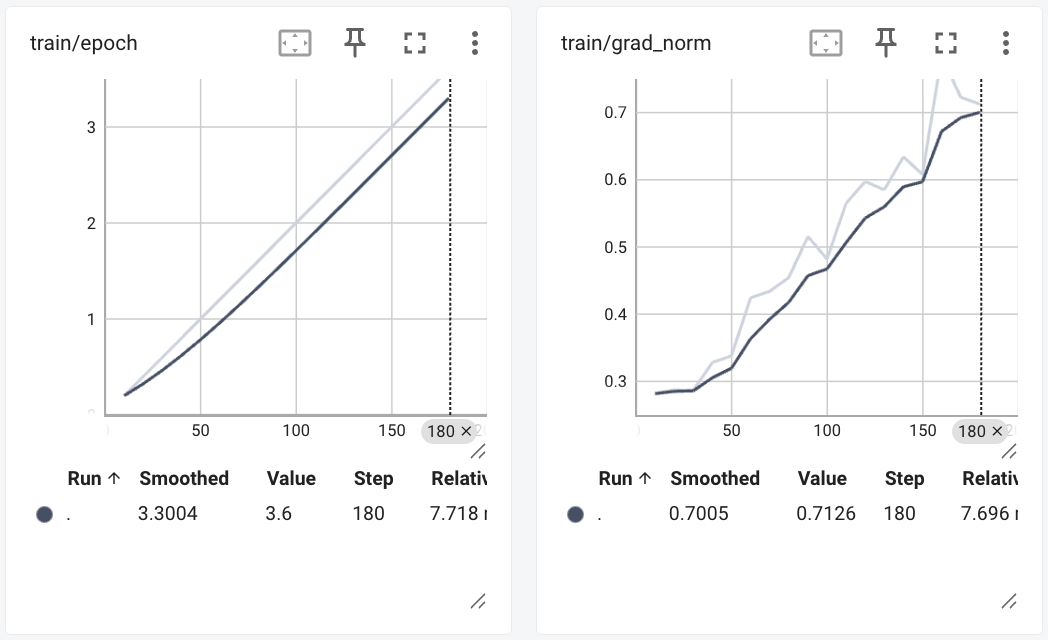}
\DeclareGraphicsExtensions.
\caption{Training progression metrics: epoch count (left) showing linear progression through approximately 3.6 epochs, and gradient norm (right) increasing gradually from 0.28 to 0.71, indicating stable optimization without gradient explosion.}
\label{fig:mental-grad-norm}
\end{figure}

The gradient norm, depicted in Figure~\ref{fig:mental-grad-norm}, increased gradually from approximately 0.28 to 0.71 throughout training. This controlled increase indicates stable gradient flow without exploding gradients, validating the effectiveness of the chosen hyperparameters and LoRA configuration~\cite{llama-recipe}.

\subsubsection{Resource Utilization}

Training was completed in 494.24 seconds (8.24 minutes) on a single NVIDIA A100-SXM4-40GB GPU. Table~\ref{tab:mental-resources} summarizes the computational resource utilization during fine-tuning.

\begin{table}[H]
\centering
\caption{Training Resource Utilization for Psychiatric Assessment Model}
\label{tab:mental-resources}
\begin{tabular}{ll}
\toprule
\textbf{Metric} & \textbf{Value} \\
\midrule
Training Duration & 494.24 seconds (8.24 minutes) \\
Hardware & NVIDIA A100-SXM4-40GB \\
GPU Maximum Memory & 39.557 GB \\
Initial Memory Reserved & 6.725 GB \\
Peak Memory Reserved & 7.789 GB \\
Memory for LoRA Training & 1.064 GB \\
Training Steps Completed & 180 (early stopping) \\
Number of Epochs & $\approx$3.6 \\
Total FLOPs & $5.65 \times 10^{16}$ \\
\bottomrule
\end{tabular}
\end{table}

The peak reserved memory during training was 7.789~GB, with the initial model loading consuming 6.725~GB. The additional memory required for LoRA fine-tuning was only 1.064~GB, representing approximately 13.7\% overhead.

\begin{figure}[H]
\centering{}
\includegraphics[width=5.2in]{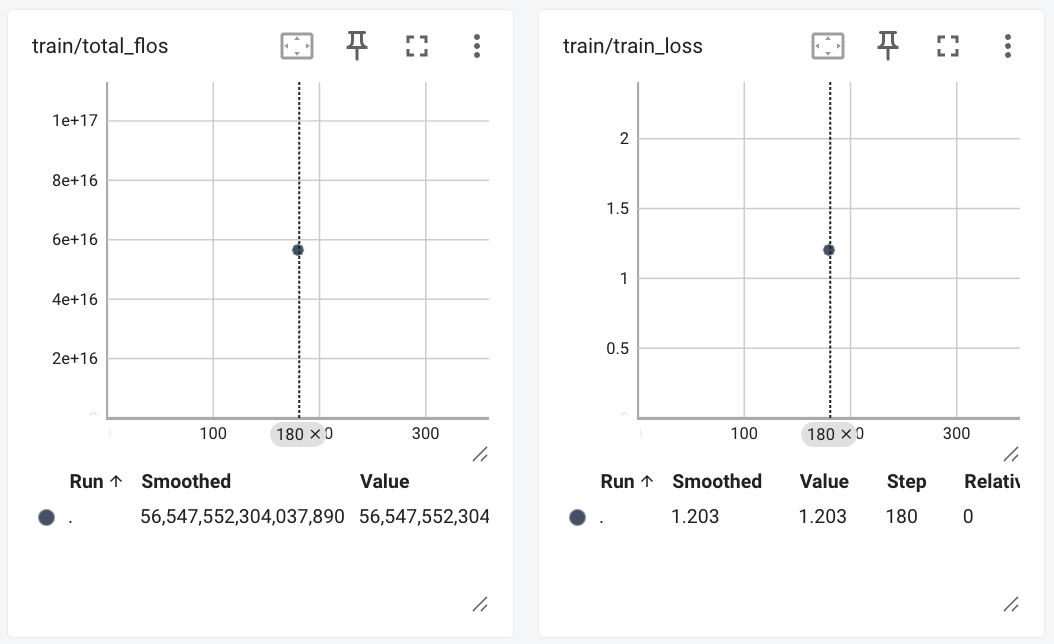}
\DeclareGraphicsExtensions.
\caption{Computational efficiency metrics: total floating-point operations (left) reaching $5.65 \times 10^{16}$ FLOPs, and cumulative training loss (right) at 1.203, demonstrating the computational cost of parameter-efficient fine-tuning.}
\label{fig:mental-flops}
\end{figure}

Figure~\ref{fig:mental-flops} illustrates the total floating-point operations (FLOPs) and cumulative training loss. The fine-tuning process required approximately $5.65 \times 10^{16}$ FLOPs, demonstrating the computational efficiency of LoRA-based adaptation compared to full model fine-tuning, which would require significantly higher computational resources~\cite{lora, wedagpt}.

These efficiency metrics demonstrate that parameter-efficient fine-tuning enables adaptation of large language models for specialized psychiatric assessment tasks using accessible cloud-based infrastructure.

\subsubsection{Inference Performance}

\begin{figure}[H]
\centering{}
\includegraphics[width=5.2in]{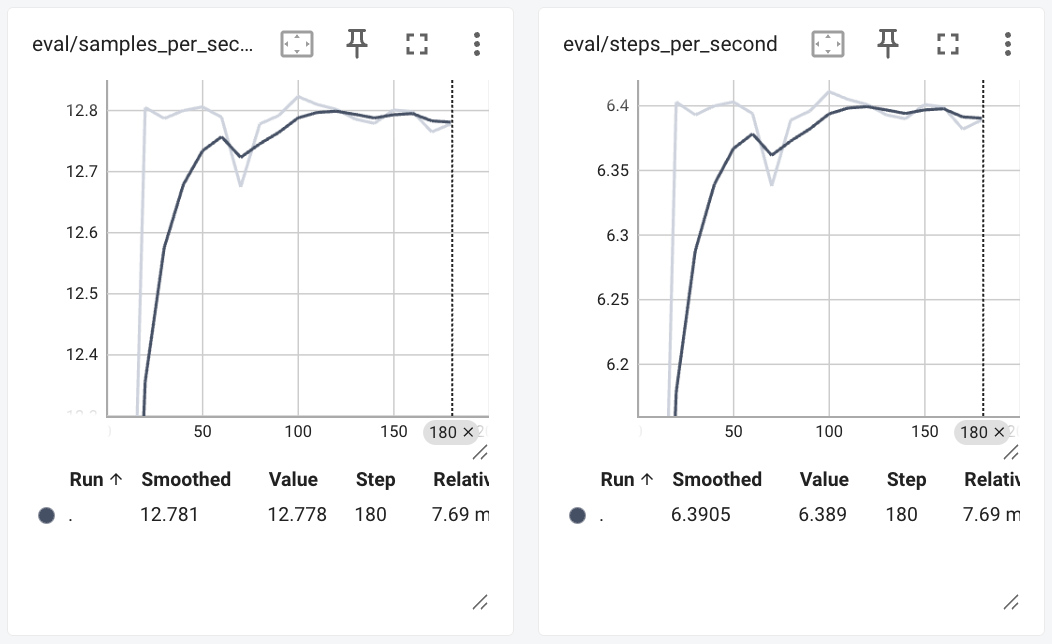}
\DeclareGraphicsExtensions.
\caption{Inference performance metrics during validation: samples per second (left) stabilizing at approximately 12.78, and steps per second (right) at approximately 6.39, indicating consistent evaluation throughput throughout training.}
\label{fig:mental-inference}
\end{figure}

Inference throughput metrics, shown in Figure~\ref{fig:mental-inference}, indicate stable evaluation performance throughout training. The model achieved approximately 12.78 samples per second and 6.39 steps per second during validation, with evaluation runtime stabilizing at approximately 3.91 seconds per evaluation batch.

\subsection{Predictive Performance of Fine-Tuned LLMs}

Following the fine-tuning phase, we evaluated the predictive performance of the fine-tuned LLM consortium in the context of structured psychological assessment support. The evaluation compared outputs generated by baseline (pre-trained) LLMs and their fine-tuned counterparts against reference annotations derived from psychiatrist–patient conversation data. These annotations, including DSM-5 diagnostic codes, were used solely as structured labels for assessing alignment and consistency, rather than as indicators of autonomous diagnostic capability~\cite{dsm-5-criteria}.

\paragraph{Quantitative Assessment Alignment.}
To assess predictive alignment quantitatively, the fine-tuned model was evaluated on a held-out test set comprising 50 assessment samples. Table~\ref{tab:mental-accuracy} summarizes the alignment accuracy across five psychiatric reference categories.

\begin{table}[H]
\centering
\caption{Assessment Label Alignment Accuracy by Condition}
\label{tab:mental-accuracy}
\begin{tabular}{lccc}
\toprule
\textbf{Condition} & \textbf{Correct} & \textbf{Total} & \textbf{Accuracy} \\
\midrule
Generalized Anxiety Disorder & 13 & 13 & 100.0\% \\
Bipolar I Disorder & 14 & 14 & 100.0\% \\
Panic Disorder & 8 & 8 & 100.0\% \\
Major Depressive Disorder & 5 & 8 & 62.5\% \\
Post-Traumatic Stress Disorder & 4 & 7 & 57.1\% \\
\midrule
\textbf{Overall} & \textbf{44} & \textbf{50} & \textbf{88.0\%} \\
\bottomrule
\end{tabular}
\end{table}

The model achieved perfect alignment for Generalized Anxiety Disorder, Bipolar I Disorder, and Panic Disorder, indicating strong pattern recognition for these symptom clusters. In contrast, performance on Major Depressive Disorder and Post-Traumatic Stress Disorder was limited, reflecting the higher degree of symptom overlap, contextual dependence, and presentation variability associated with these conditions. These results highlight both the potential and limitations of LLM-based assessment support and reinforce the importance of human-in-the-loop oversight within the Train-the-Trainers framework.

\paragraph{Qualitative Assessment Outputs}
In addition to quantitative evaluation, we examined representative assessment outputs to assess improvements in structure, clarity, and interpretability following fine-tuning.

\begin{figure}[H]
\centering{}
\includegraphics[width=5.2in]{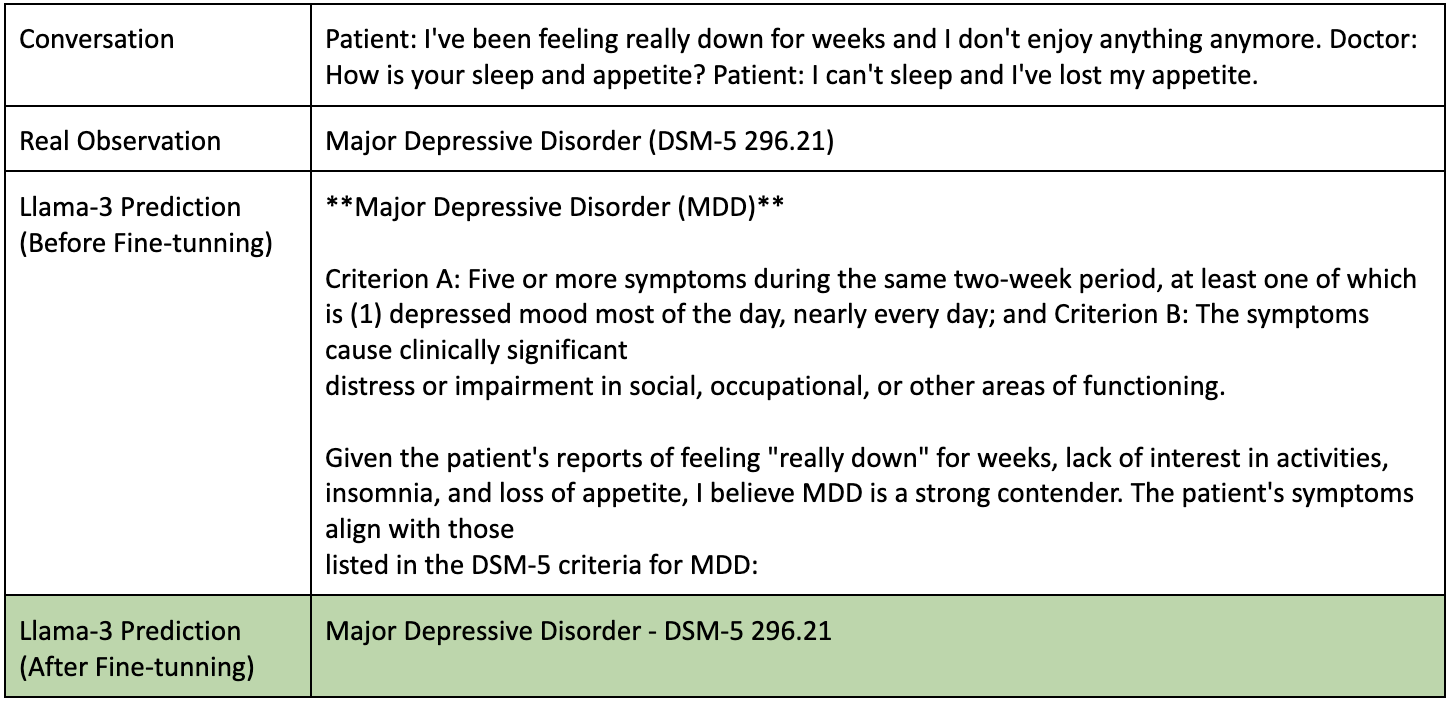}
\DeclareGraphicsExtensions.
\caption{Representative assessment output of the fine-tuned Llama-3 LLM aligned with structured reference labels (DSM-5 296.21).}
\label{prediction-llama-v1}
\end{figure}

Figure~\ref{prediction-llama-v1} illustrates a representative assessment output generated by the fine-tuned Llama-3 model. Prior to fine-tuning, Llama-3 produced verbose and loosely structured responses that referenced relevant symptoms but did not consistently align observations with standardized assessment categories. After fine-tuning, the model generated concise and structured assessment summaries that aligned more reliably with the corresponding reference labels, reflecting improved reasoning over symptom narratives and contextual cues~\cite{llama-3}.

\begin{figure}[H]
\centering{}
\includegraphics[width=5.2in]{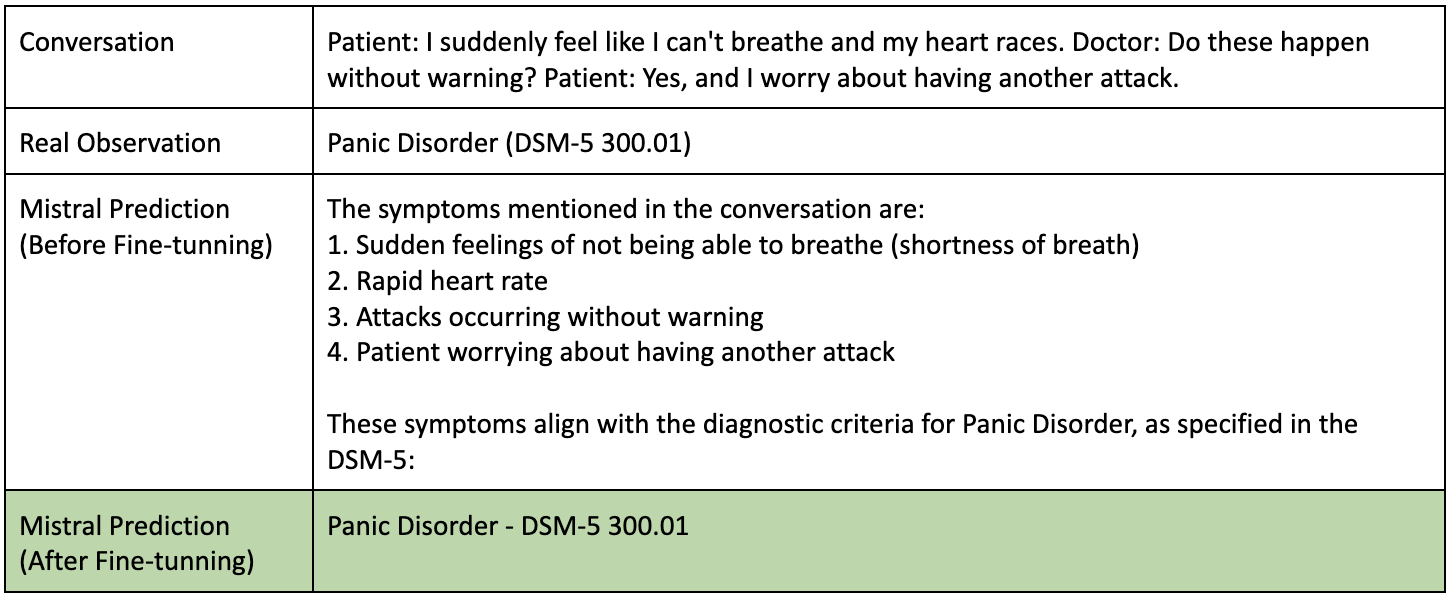}
\DeclareGraphicsExtensions.
\caption{Representative assessment output of the fine-tuned Mistral LLM aligned with structured reference labels (DSM-5 300.01).}
\label{prediction-mistral-v2}
\end{figure}

Figure~\ref{prediction-mistral-v2} presents a representative assessment output produced by the fine-tuned Mistral model~\cite{pixtral}. While the baseline model identified salient symptom patterns, its outputs were often verbose and interpretive. Fine-tuning resulted in more precise, consistently structured assessment representations, reducing ambiguity and improving alignment with the reference framework.

\begin{figure}[H]
\centering{}
\includegraphics[width=5.2in]{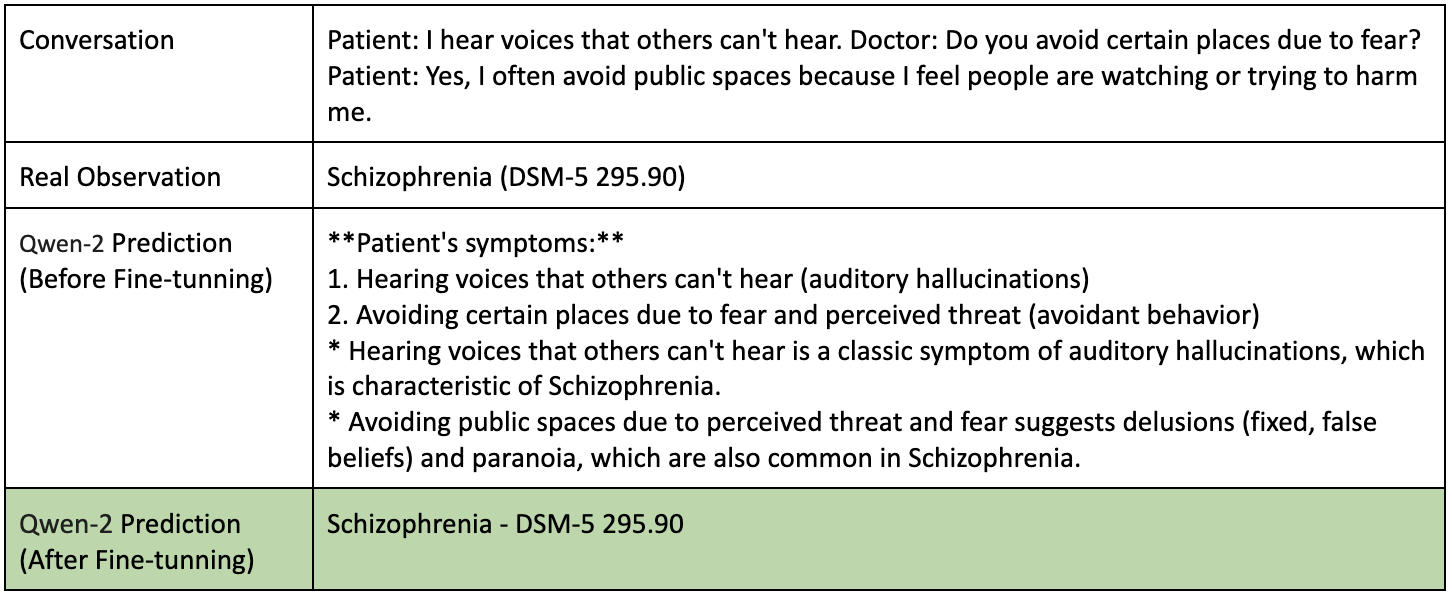}
\DeclareGraphicsExtensions.
\caption{Representative assessment output of the fine-tuned Qwen2 LLM aligned with structured reference labels (DSM-5 295.90).}
\label{prediction-qwen-v2}
\end{figure}

Figure~\ref{prediction-qwen-v2} shows a representative assessment output generated by the fine-tuned Qwen2 model~\cite{qwen2, on-device-qwen2}. Prior to fine-tuning, Qwen2 exhibited partial alignment with reference categories but occasionally lacked clarity and consistency. After fine-tuning, the model demonstrated improved mapping between observed symptom clusters and structured assessment labels, yielding outputs that were more concise, interpretable, and aligned with the reference framework.

Across all evaluated models, fine-tuning consistently improved predictive alignment, output structure, and interpretability. Similar qualitative improvements were observed across multiple assessment scenarios and reference conditions. Together, these results demonstrate that task-specific fine-tuning substantially enhances the suitability of LLMs as decision-support components within the Assessment Agent, supporting reliable and interpretable human-led assessment and escalation decisions under operational constraints~\cite{llm-reasoning}.

\subsection{Evaluation of Consensus-Driven Reasoning}

This evaluation examines the performance of the reasoning-oriented LLM (GPT-OSS) in synthesizing and validating assessment outputs generated independently by multiple fine-tuned LLMs within the consortium~\cite{llm-reasoning}. The objective is to assess the system’s ability to perform consensus-driven reasoning, reconciling diverse model perspectives into a coherent, interpretable, and reliable assessment representation suitable for human decision support.

\begin{figure}[H]
\centering{}
\includegraphics[width=5.2in]{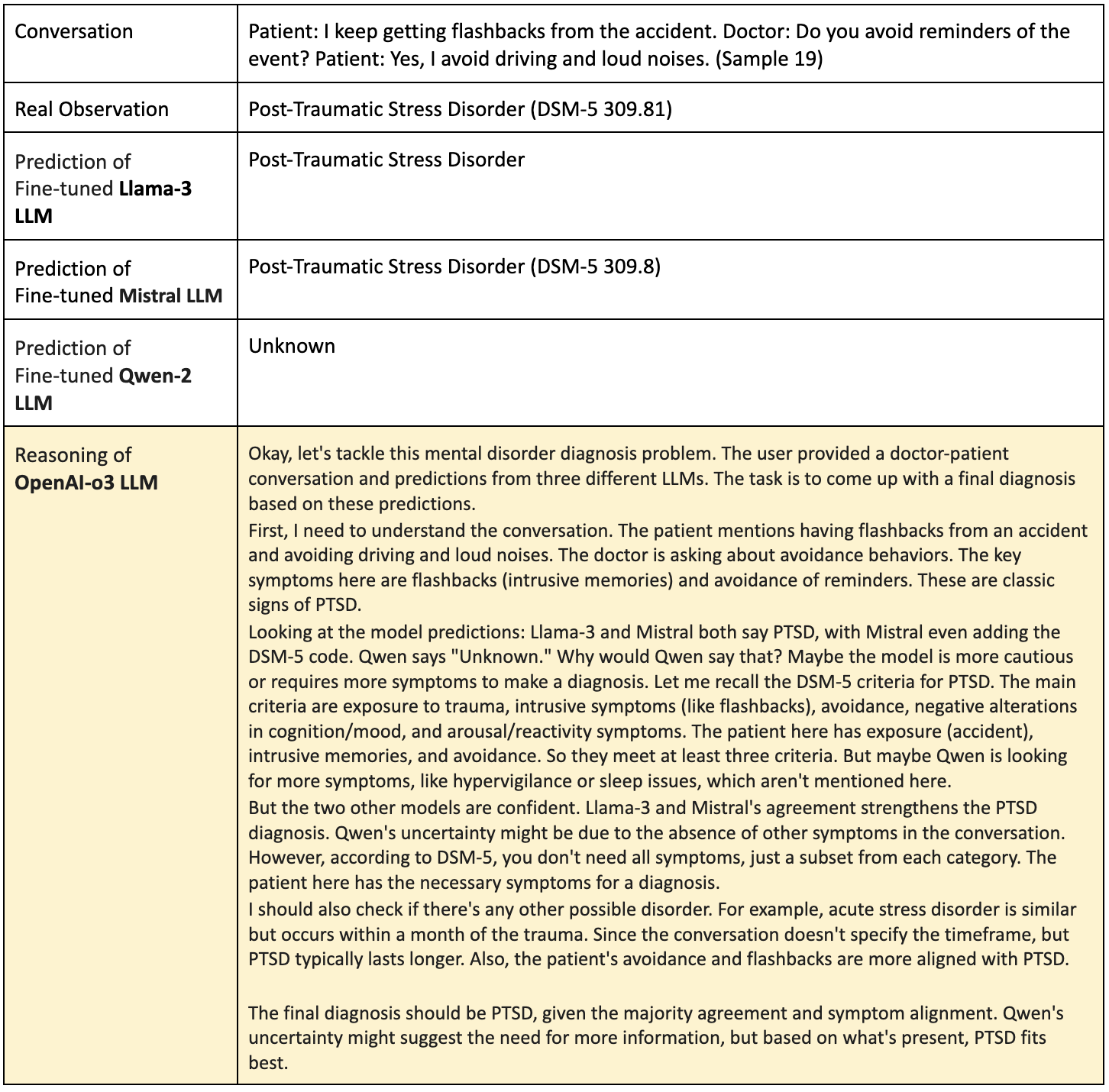}
\DeclareGraphicsExtensions.
\caption{Consensus-driven assessment synthesis produced by the reasoning-oriented LLM through multi-model evaluation and reconciliation.}
\label{prediction-o3}
\end{figure}

In the evaluated workflow, individual fine-tuned LLMs generate structured assessment outputs based on the same input context, reflecting different reasoning emphases and interpretations. The reasoning-oriented LLM then evaluates these outputs, identifies areas of agreement and disagreement, and synthesizes a consolidated assessment accompanied by an explicit rationale. Figure~\ref{prediction-o3} presents an illustrative example comparing the independent assessment outputs from the Llama-3, Mistral, and Qwen2 models with the final synthesized output produced by the reasoning LLM~\cite{towards-rai-xai}.

The results demonstrate that the reasoning-oriented LLM effectively resolves inconsistencies across model outputs by prioritizing coherence, contextual relevance, and alignment with the reference assessment framework. Rather than simply selecting a majority outcome, the reasoning model evaluates the supporting evidence provided by each model and produces a consolidated representation that reflects the most consistent and contextually grounded interpretation~\cite{nurolense}. This process reduces single-model bias and mitigates the impact of spurious or overconfident predictions.

From a responsible and explainable AI perspective, the consensus-driven reasoning layer provides several key benefits. First, it introduces transparency by exposing intermediate model outputs and the rationale underlying the final synthesized assessment. Second, it improves robustness by ensuring that no single model’s prediction dominates the outcome. Third, it preserves human oversight by presenting the peer trainer with an interpretable assessment summary rather than an opaque or authoritative decision.

Importantly, the reasoning LLM operates strictly in an advisory capacity and does not perform autonomous diagnosis or intervention. Its role is limited to synthesizing assessment-level information to support human judgment within the Train-the-Trainers framework. The evaluation results indicate that incorporating a dedicated reasoning-oriented LLM as a consensus layer significantly enhances the reliability, interpretability, and trustworthiness of agentic AI decision support in sensitive and high-stakes environments~\cite{agentic-ai-workflow-patterns}.

\section{Related Work}

Recent research has demonstrated the growing potential of large language models (LLMs) and AI-driven decision-support systems across a range of healthcare domains, including clinical reasoning, medical question answering, and diagnostic assistance. Systems such as Med-PaLM~\cite{med-palm}, Me-LLaMA~\cite{me-llama}, and LLM-based differential diagnosis frameworks~\cite{llm-ddx} have shown that domain-adapted language models can achieve strong performance on medical benchmarks and assist clinicians in structured reasoning tasks. In mental health specifically, prior work has explored AI-assisted psychiatric assessment and diagnostic support using fine-tuned LLMs and hybrid reasoning pipelines~\cite{CDSS}. Other systems, such as DrHouse~\cite{drhouse}, extend LLM-based decision support by incorporating multimodal inputs and iterative reasoning for primary care triage. However, most existing approaches are designed for civilian clinical settings, rely on centralized infrastructure, and assume direct clinician interaction. They typically emphasize autonomous or semi-autonomous diagnostic output rather than supporting peer-led intervention under operational constraints.

In parallel, advances in agentic AI and multi-agent systems have introduced new paradigms for automating complex workflows through coordinated reasoning, tool use, and iterative decision-making~\cite{agentsway}. Recent studies have shown that agentic systems can leverage ensembles or consortia of LLMs combined with dedicated reasoning models to improve robustness, reduce single-model bias, and enhance interpretability~\cite{reasoning-llms, wedagpt}. Consensus-driven reasoning approaches, including multi-model synthesis and structured deliberation, have proven particularly effective in high-uncertainty tasks~\cite{llm-ddx}. Despite these advances, most agentic AI systems prioritize autonomy and task completion, with limited emphasis on strict human-in-the-loop control, explainability, or deployment in high-stakes and resource-constrained environments such as military operations.

In contrast, the proposed Train-the-Trainers framework uniquely integrates peer-based mental health support with a human-supervised, agentic AI architecture tailored for forward-deployed and austere settings. Rather than automating diagnosis or care delivery, the system augments recovered soldiers acting as peer trainers with structured decision support across assessment, peer intervention guidance, operational reasoning, and escalation. The use of a consensus-driven LLM consortium and a reasoning-oriented model enables transparent and explainable recommendations while preserving human authority and ethical boundaries. To the best of our knowledge, this work represents the first application of agentic AI to peer-led mental health support in battlefield environments, bridging gaps between military mental health practice, responsible AI design, and deployable multi-agent systems.

Table~\ref{tab:comparison-frameworks} summarizes key differences between prior AI-based healthcare and mental health systems and the proposed Train-the-Trainers framework.

\begin{table*}[!htb]
\centering
\caption{Comparison of AI-Based Healthcare and Mental Health Support Frameworks}
\label{tab:comparison-frameworks}
\begin{adjustbox}{width=\textwidth}
\begin{tabular}{lccccccc}
\toprule
\textbf{Framework} &
\textbf{Domain} &
\textbf{Agentic AI} &
\textbf{LLM Fine-Tuning} &
\textbf{Supported LLMs} &
\textbf{Human-in-the-Loop} &
\textbf{Responsible AI} &
\textbf{Explainable AI} \\
\midrule

\makecell{Train-the-Trainers\\(This Work)} &
Battlefield Mental Health &
\cmark &
\cmark &
\makecell{LLaMA-3, Mistral, Qwen \\ OpenAI-gpt-oss} &
\cmark &
\cmark &
\cmark \\

Med-PaLM~\cite{med-palm} &
General Medicine &
\xmark &
\cmark &
PaLM &
\cmark &
\cmark &
\xmark \\

LLM for DDx~\cite{llm-ddx} &
General Medicine &
\xmark &
\cmark &
Multiple (unspecified) &
\cmark &
\xmark &
\cmark \\

Me-LLaMA~\cite{me-llama} &
Biomedical NLP &
\xmark &
\cmark &
LLaMA &
\cmark &
\xmark &
\xmark \\

DrHouse~\cite{drhouse} &
Primary Care &
\cmark &
\cmark &
Multiple (unspecified) &
\cmark &
\cmark &
\cmark \\

CDSS~\cite{CDSS} &
Mental Health &
\xmark &
\xmark &
GPT-4 &
\cmark &
\cmark &
\cmark \\

Weda-GPT~\cite{wedagpt} &
Indigenous Medicine &
\xmark &
\cmark &
LLaMA-3 &
\cmark &
\cmark &
\cmark \\

\bottomrule
\end{tabular}
\end{adjustbox}
\end{table*}

\section{Conclusions and Future Work}

This paper presented a Train-the-Trainers framework augmented by an agentic AI platform to support peer-based mental health assessment and intervention in forward-deployed and resource-constrained environments. By positioning recovered soldiers as peer trainers and equipping them with a human-supervised, multi-agent AI system, the proposed approach addresses critical gaps in timely psychological support where access to professional clinicians is limited. The framework emphasizes early assessment, structured peer support, informed escalation, and continuity of care while preserving clear boundaries between decision support and clinical authority.

We described the design and implementation of a functional prototype that integrates multiple specialized AI agents under a strict human-in-the-loop orchestration model. Central to the system is an Assessment Agent implemented using a consortium of fine-tuned large language models and a reasoning-oriented LLM that performs consensus-driven synthesis. The evaluation focused on the technical performance of this assessment pipeline, demonstrating that task-specific fine-tuning improves the consistency and structure of assessment outputs, while the consensus-driven reasoning layer enhances robustness, interpretability, and reliability. Importantly, all AI components operate in an advisory capacity, supporting human judgment rather than performing autonomous diagnosis or intervention.

The results indicate that agentic AI, when carefully designed with human oversight, explainability, and operational constraints in mind, can function as a force multiplier for peer-based mental health support in austere environments. The proposed framework is compatible with air-gapped deployment, emphasizes transparency and accountability, and aligns with responsible AI principles required for sensitive military and humanitarian applications.

Several directions for future work emerge from this study. First, broader evaluation using field-generated data and longitudinal studies is needed to assess system performance under real operational conditions and evolving stress patterns. Second, future work will explore the integration of additional contextual signals, such as physiological or environmental indicators, while maintaining privacy and ethical safeguards. Third, user-centered evaluations involving peer trainers can provide insights into usability, trust, and cognitive load during high-stress interactions. Finally, extending the framework to support coordination with remote clinicians, secure telehealth interfaces, and cross-unit learning mechanisms represents a promising pathway for scaling the Train-the-Trainers approach across diverse operational contexts.

Overall, this work demonstrates the feasibility and potential of combining peer-based intervention models with consensus-driven, human-supervised agentic AI. By prioritizing responsible design, explainability, and human control, the proposed system offers a practical foundation for enhancing mental health support in environments where timely care is most difficult to deliver.

% \section*{Acknowledgements}

% This work was supported in part by the DoD Center of Excellence in AI and Machine Learning (CoE-AIML) under Contract Number W911NF-20-2-0277 with the U.S. Army Research Laboratory. 

\bibliographystyle{IEEEtran}
\bibliography{reference}

\end{document}